\documentclass[12pt]{article}

\usepackage{amsmath,amsfonts,amssymb}

\def\ttau{\tilde{\tau}}

\def\be{\begin{equation}}

\def\ee{\end{equation}}

\def\bea{\begin{eqnarray}}

\def\eea{\end{eqnarray}}

\def\mH{\mathcal{H}}

\def\ba{\mathbf{a}}

\def\bh{\bar{h}}

\newcommand{\bT}{\mathbf{T}}

\newcommand{\mL}{\mathcal{L}}

\def\pb #1{\left\{#1\right\}}

\begin{document}

	\begin{titlepage}

		\vskip 0.4 cm
		
		\begin{center}
			{\Large{ \bf Hamiltonian For String in Newton-Cartan Background 			}}
			
			\vspace{1em} Josef Kluso\v{n}$\,^1$, 
			\footnote{Email address:
				klu@physics.muni.cz }\\
			\vspace{1em} $^1$\textit{Department of Theoretical Physics and
				Astrophysics, Faculty
				of Science,\\
				Masaryk University, Kotl\'a\v{r}sk\'a 2, 611 37, Brno, Czech Republic}\\

			%
			%
			
			\vskip 0.8cm
			
		\end{center}

		\begin{abstract}
This paper is devoted to the construction
of the Hamiltonian for non-relativistic string in the Newton-Cartan
background. We start with the Hamiltonian for relativistic string in
general background. Then we perform limiting procedure on the metric
that leads to Newton-Cartan background. We determine  constraint
structure for non-relativistic string and show that these
constraints are the first class constraints. Then we determine
corresponding Lagrangian and discuss its properties. 		
			
		\end{abstract}
		
		\bigskip
		
	\end{titlepage}
	
	\newpage

\section{Introduction}
Recently, non-Lorentzian geometry has gained interest in theoretical
physics community from many reasons. Firstly, today it is well known
that strong correlated systems  in condensed matter can be
successfully described with the help of non-relativistic holography
\cite{Christensen:2013lma,Christensen:2013rfa,Hartong:2014oma} , for
review see for example \cite{Hartnoll:2016apf}. This duality is
based on the idea that the strongly coupled theory on the boundary
can be described by string theory in the bulk. Further, when the
curvature of the space-time is small we can use the classical
gravity instead of the  full string theory machinery. In case of
non-relativistic holography the situation is even more interesting
since we have basically two possibilities: Either we  use Einstein
metric with non-relativistic isometries
\cite{Son:2008ye,Balasubramanian:2008dm,Herzog:2008wg} or we
introduce non-relativistic gravities in the bulk
\cite{Son:2013rqa,Janiszewski:2012nb}, like Newton-Cartan gravity
\cite{Cartan:1923zea} \footnote{For some recent works, see
\cite{Bergshoeff:2017dqq,Bergshoeff:2016lwr,Afshar:2015aku,
Bergshoeff:2015ija,Bergshoeff:2015uaa,Bergshoeff:2014uea,
Andringa:2010it,Hartong:2015zia,Hartong:2016yrf,Bergshoeff:2017btm,Bergshoeff:2017dqq,
Grosvenor:2017dfs,Jensen:2014aia,Jensen:2014ama,Jensen:2014wha}.}
 or Ho\v{r}ava gravity \cite{Horava:2009uw}. It is also very instructive
 to analyze extended objects in Newton-Cartan theory
\cite{Andringa:2012uz,Harmark:2017rpg} \footnote{For the analysis of
point particles in this background, see
\cite{Barducci:2017mse,Kluson:2017pzr}.}. In \cite{Andringa:2012uz}
the action for non-relativistic string in Newton-Cartan background
was proposed that has many interesting properties. For example, in
was argued in \cite{Andringa:2012uz} that in order to define
correctly an action for non-relativistic string in Newton-Cartan
background two longitudinal directions have to be selected and hence
we obtain more general form of the Newton-Cartan geometry. The
canonical analysis of this string was performed recently in
\cite{Kluson:2017abm}. During this analysis we met an obstacle which
was an impossibility to derive Hamiltonian constraint for the string
with non-zero gauge field $m_\mu^{ \ a}$ that will be defined in the
next section. For that reason we were forced to restrict to the case
of zero gauge field $m_\mu^{ \ a}$ and then we were able to
determine canonical structure of the non-relativistic string in
Newton-Cartan background. In the same way we proceeded with the case
of non-relativistic p-brane. We defined it using the limiting
procedure introduced in \cite{Bergshoeff:2015uaa}. We again found
corresponding action for non-relativistic p-brane in Newton-Cartan
background and determined   canonical structure for this theory  on
condition that the gauge field $m_\mu^{ \ a}$ is zero.

The fact that in our previous work we considered the situation when
the gauge field $m_\mu^{ \ a}$  vanishes   is rather unsatisfactory
since this field is crucial for the invariance of the theory under
Milne boost. It would be nice to develop full canonical formalism
where this field is non-zero. We suggested in the conclusion of our
previous paper \cite{Kluson:2017abm} that one way how to proceed is
to start with the Hamiltonian for the string in general background
and then perform the limiting procedure when we generalize the
approach introduced in \cite{Bergshoeff:2015uaa} to the case of two
longitudinal directions. Exactly this is the goal of our paper.  We
start with the Hamiltonian for relativistic string in general
background, introduce relativistic vierbeins  and NSNS two form that
are functions of fields that define Newton-Cartan background. These
fields also depend on the free parameter that goes to infinity when
we define Newton-Cartan gravity \cite{Bergshoeff:2015uaa}.  As a
result we will be able to find corresponding Hamiltonian for the
string in Newton-Cartan background. However this is not certainly
the end of the story since we have to perform consistency checks of
this proposal. Explicitly, we have to show that constraints, that
define this theory, are the first class constraints. It turns out
that this is a non-trivial task due to the complicated form of the
Hamiltonian. Secondly, we would like to find  Lagrangian for this
non-relativistic string and investigate how it is related to the
Lagrangian density proposed in \cite{Andringa:2012uz}. To do this we
carefully examine an invariance of the Hamiltonian constraint  under
generalized Milne boost. We show that the Hamiltonian constraint can
be rewritten with the help of variables that are manifestly
invariant under Milne transformation so that Hamiltonian is
invariant too. Then we can proceed to the analysis of corresponding
Lagrangian. As a warm up we consider the case of the
non-relativistic string in flat background. We show that there is a
crucial difference between inverse Legendre transformation in case
of the relativistic string and non-relativistic one. Explicitly, we
show that in case of non-relativistic string the Lagrange
multipliers corresponding to Hamiltonian and spatial diffeomorphism
constraints are determined by projections of the equations of motion
for $x^\mu$ to longitudinal directions instead of their equations of
motion. Then we will be able to find Lagrangian that agrees with the
Lagrangian found in \cite{Andringa:2012uz}. Further we proceed to
the most general case of the non-relativistic string in
Newton-Cartan background where the analysis is much more
complicated. Despite of this fact we find Lagrangian form of the
non-relativistic string in Newton-Cartan background which is
manifestly diffeomorphism invariant and which agrees with the Lagrangian
density proposed in \cite{Andringa:2012uz}.


This paper is organized as follows. In the next section (\ref{second}) we introduce canonical form of the relativistic string action and perform limiting procedure that leads to the Hamiltonian for non-relativistic string in
Newton-Cartan background and determine Poisson algebra of constraints. In section (\ref{third}) we find the Lagrangian for non-relativistic string in Newton-Cartan background. Finally in conclusion (\ref{fourth}) we outline our results and suggest possible extension of this work.

 \section{Canonical Formulation of Non-relativistic String in Newton-Cartan  Background}\label{second}
We start with the  Nambu-Gotto form of the action for relativistic string in  general background
 \begin{equation}\label{funstringact}
 S=-\ttau_F \int d\tau d\sigma\sqrt{-\det (E_\mu^{ \ A}E_\nu^{ \ B}\eta_{AB}
    \partial_\alpha x^\mu\partial_\beta x^\nu)}+\ttau_F
 \int d\tau d\sigma B_{\mu\nu}\partial_\tau x^\mu \partial_\sigma x^\nu \ ,
 \end{equation}
 where $E_\mu^{ \ A}$ is $d-$dimensional vierbein so that the metric components have the form
 \begin{equation}
 G_{\mu\nu}=E_\mu^{ \ A}E_\nu^{ \ B}\eta_{AB} \ , \eta_{AB}=\mathrm{diag}(-1,\dots,1) \ .
 \end{equation}
 Note that the metric inverse $G^{\mu\nu}$ is defined with the help of the inverse vierbein $E^\mu_{ \ B}$ that obeys the relation
 \begin{equation}
 E_\mu^{ \ A}E^\mu_{ \ B}=\delta^A_{B} \  ,  \quad E_\mu^{ \ A}E^\nu_{ \ A}=
 \delta^\mu_{\nu} \ .
 \end{equation}
 Further, $B_{\mu\nu}$ is NSNS two form field. Finally $x^\mu \ ,\mu=0,\dots,d-1$ are embedding coordinates of the string where the two dimensional world-sheet is parameterized by  $\sigma^\alpha\equiv(\tau,\sigma)$ and $\ttau_F$ is the string tension that could be eventually rescaled when we define non-relativistic string.

 Our goal is to find Hamiltonian  non-relativistic string in Newton-Cartan background with the help of the following procedure. As the first step we determine
Hamiltonian  from the action (\ref{funstringact}).
Explicitly, from (\ref{funstringact}) we find following conjugate momenta
\begin{equation}\label{pmu}
 p_\mu=-\ttau_F E_\mu^{ \ A}E_\nu^{ \ B}\eta_{AB}\partial_\beta x^\nu
 g^{\beta\tau}\sqrt{-\det g_{\alpha\beta}}+\ttau_F                                                                                                                                                                      B_{\mu\nu}\partial_\sigma x^\nu \ ,
 \end{equation}
 where
 \begin{equation}
 g_{\alpha\beta}\equiv G_{\mu\nu}\partial_\alpha x^\mu \partial_\beta x^\nu \ , \quad  g^{\alpha\beta}g_{\beta\gamma}=\delta^\alpha_\gamma \ .
 \end{equation}
 Using (\ref{pmu}) we immediately find that the bare Hamiltonian
 $H_B=\int d\sigma (p_\mu \partial_\tau x^\mu-\mL)$ is zero
 while  we have following two primary   constraints
 \begin{eqnarray}
  \mH_\tau &\equiv& (p_\mu-\ttau_F B_{\mu\rho}\partial_\sigma x^\rho)
 E^\mu_{ \ A}E^\nu_{ \ B}\eta^{AB}(p_\nu-\ttau_F B_{\nu\sigma}
 \partial_\sigma x^\sigma)+ \nonumber \\
 &+& \ttau_F^2 \partial_\sigma x^\mu E_\mu^{ \ A}
 \eta_{AB}E_\nu^{ \ B}\partial_\sigma x^\nu\approx 0 \ , \quad
\mH_\sigma \equiv p_\mu \partial_\sigma x^\mu \approx 0 \ . \nonumber \\
 \end{eqnarray}

Now we are ready to find Hamiltonian for the string in Newton-Cartan background with the help
of the  non-relativistic limit of relativistic vierbein $E_\mu^{ \ A}$ \cite{Bergshoeff:2015uaa}. However as we argued
in our recent paper \cite{Kluson:2017abm} in order to find correct non-relativistic
limit we have to introduce the generalization of Newton-Cartan gravity following
\cite{Andringa:2012uz}. Explicitly,
we  split target-space indices $A$ into $A=(a',a)$ where now $a=0,1$ and $a'=2,\dots,d-1$. Then we introduce $\tau_\mu^{ \ a}$ so that we write
 \begin{equation}
  \tau_{\mu\nu}=\tau_\mu^{ \ a}\tau_\nu^{ \ b}
 \eta_{ab} \ , \quad  a,b=0,1 \ .
 \end{equation}
 In the same way we introduce vierbein $e_\mu^{ \ a'}, a'=2,\dots,d-1$ and also  introduce gauge field  $m_\mu^{ \ a}$. The $\tau_\mu^{ \ a}$ can be interpreted as the gauge fields of the longitudinal translations while $e_\mu^{ \ a'}$  as the gauge fields of the transverse translations
 \cite{Andringa:2012uz}. Then we can also introduce their inverses with respect to their longitudinal and transverse translations
 \begin{eqnarray}
 e_\mu^{ \ a'}e^\mu_{ \ b'}=\delta^{a'}_{b'} \ ,  \quad
 e_\mu^{ \ a'}e^\nu_{ \ a'}=\delta_\mu^\nu-\tau_\mu^{ \ a}
 \tau^\nu_{ \ a} \ , \quad \tau^\mu_{ \ a}\tau_\mu^{ \ b}=\delta_a^b \ , \quad
 \tau^\mu_{ \ a}e_\mu^{ \ a'}=0 \ , \quad
 \tau_\mu^{ \ a}e^\mu_{ \ a'}=0 \ . \nonumber \\
 \end{eqnarray}
 Now we are ready
to introduce following parameterization of the  vierbein  $E_\mu^{ \ A}$
 \cite{Bergshoeff:2015uaa}
 \begin{equation}\label{relvier}
 E_\mu^{ \ a}=\omega \tau_\mu^{ \ a}+\frac{1}{2\omega}m_\mu^{ \ a} \ , \quad
 E_\mu^{ \ a'} =e_\mu^{ \ a'} \ ,
 \end{equation}
 where $\omega$ is a free parameter that we take to infinity when we define non-relativistic limit. Note that the inverse vierbein to (\ref{relvier}) has the form  (up to terms of order $\omega^{-3}$)
 \begin{equation}\label{relvierinv}
 E^\mu_{ \ a}=\frac{1}{\omega}\tau^\mu_{ \ a}-\frac{1}{2\omega^3}\tau^\mu_{ \ b}m_\rho^{ \ b}
 \tau^\rho_{ \ a} \ , \quad E^\mu_{ \ a'}=e^\mu_{ \ a'}
 -\frac{1}{2\omega^2} \tau^\mu_{ \ a}m_\rho^{ \ a}e^\rho_{ \ a'} \ .
 \end{equation}
 Then with the help of (\ref{relvier}) and (\ref{relvierinv}) we obtain following form of the metric $G_{\mu\nu}$ and its inverse
 \begin{eqnarray}
 G_{\mu\nu}&=&E_\mu^{ \ a}E_\nu^{ \ b}\eta_{ab}+E_\mu^{ \ a'}E_\nu^{ \ b'}\delta_{a'b'}
 =\nonumber \\
 &=&\omega^2 \tau_{\mu\nu}+h_{\mu\nu}+\frac{1}{2}\tau_\mu^{ \ a}m_\nu^{ \ b}\eta_{ab}+
 \frac{1}{2}m_\mu^{ \ a}\tau_\nu^{ \ b}\eta_{ab}+\frac{1}{4\omega^2}m_\mu^{ \ a}m_\nu^{ \ b}
 \eta_{ab} \ , \nonumber \\
 G^{\mu\nu}&=&E^\mu_{ \ a}E^\nu_{ \ b}\eta^{ab}+E^\mu_{ \ a'}E^\nu_{ \ b'}\delta^{a' b'}=
 \nonumber \\
&=& \frac{1}{\omega^2}\tau^{\mu\nu}+h^{\mu\nu}
 -\frac{1}{2\omega^2}(\tau^\nu_{ \ b}m_\rho^{ \ b}h^{\rho\mu}
 +\tau^\mu_{ \ b}m_\rho^{ \ b}h^{\rho\nu})
 -\nonumber \\
&-&\frac{1}{2\omega^4}
 (\tau^\mu_{ \ c}m^c_{ \ \rho}\tau^{\rho\nu}+
 \tau^\nu_{ \ d}m^d_{ \ \rho}\tau^{\rho\mu})
 +\frac{1}{4\omega^4}\tau^\mu_{ \ a}m_\rho^{ \ a}
 h^{\rho\sigma}\tau^\nu_{ \ b}m_\sigma^{\ b}+O(\omega^{-6}) \ , \nonumber \\
 \end{eqnarray}
 where
 \begin{equation}
 h^{\mu\nu}=e^\mu_{ \ a'}e^\nu_{ \ b'}\delta^{a'b'} \ , \quad
 h_{\mu\nu}=e_\mu^{ \ a'}e_\nu^{ \ b'}\delta_{a'b'} \ , \quad
 \tau^{\mu\nu}=\tau^\mu_{ \ a}\tau^\nu_{ \ b}\eta^{ab} \ .
 \end{equation}
 As the next step we have to introduce an appropriate parameterization of NSNS two form. We suggested in \cite{Kluson:2017abm}
that it is natural to consider following form of NSNS two form
 \begin{eqnarray}
 B_{\mu\nu}&=&\left(\omega\tau_\mu^{ \ a}-\frac{1}{2\omega}m_\mu^{ \ a}\right)\left( \omega\tau_\nu^{ \ b}-\frac{1}{2\omega}m_\nu^{ \ b}\right)\epsilon_{ab}
 =\nonumber \\
 &=&\omega^2\tau_\mu^{ \ a}\tau_\nu^{ \ b}\epsilon_{ab}-
 \frac{1}{2}\left(m_\mu^{ \ a}\tau_{\nu}^{ \ b}+
 \tau_\mu^{\  a}m_\nu^{ \ b}\right)\epsilon_{ab}+\frac{1}{4\omega^2}
 m_\mu^{ \ a}m_\nu^{ \ b}\epsilon_{ab} \ , \nonumber \\
 \end{eqnarray}
 where
 \begin{equation}
\epsilon_{ab}=-\epsilon_{ba} \ , \quad  \epsilon_{01}=1 \ .
 \nonumber \\
 \end{equation}
 With the help of this definition we easily find
 \begin{eqnarray}
 \frac{1}{\omega^2}\ttau_F^2 B_{\mu\sigma}\partial_\sigma x^\sigma \tau^{\mu\nu}
   B_{\nu\rho}
 \partial_\sigma x^\rho
 =-\omega^2\ttau_F^2 \tau_{\mu\nu}\partial_\sigma x^\mu \partial_\sigma x^\nu
 \nonumber \\
 \end{eqnarray}
 and we see that this divergent contribution to the Hamiltonian constraint
  \begin{equation}
 \frac{1}{\omega^2}\ttau_F^2 B_{\mu\sigma}\partial_\sigma x^\sigma \tau^{\mu\nu}
 B_{\nu\omega}
 \partial_\sigma x^\omega
 +\ttau_F^2\omega^2\partial_\sigma x^\mu\tau_{\mu\nu}\partial_\sigma x^\nu
 \end{equation}
 vanishes.
 Then we obtain that the Hamiltonian constraint has the form in the limit $\omega\rightarrow \infty$
 \begin{eqnarray}\label{mHtau}
 \mH_\tau
&=&p_\mu h^{\mu\nu}p_\nu -2\tau_Fp_\mu
\tau^\mu_{ \ a}\eta^{ab}\epsilon_{bc}\tau_\rho^{ \ c}\partial_\sigma x^\rho
+2\tau_F p_\mu h^{\mu\rho}m_\rho^{ \ b}\epsilon_{bd}
\tau_\rho^{ \ d}\partial_\sigma x^\rho+\nonumber \\
&+&2\tau_F^2\partial_\sigma x^\mu \tau_\mu^{ \ c}\epsilon_{cd}\tau^\nu_{ \ e}
\eta^{ed}m_\nu^{ \ a}\tau_\rho^{ \ b}\epsilon_{ab}\partial_\sigma x^\rho+
2\tau_F^2\partial_\sigma x^\mu \tau_\mu^{ \ a}\eta_{ab}m_\nu^{ \ b}\partial_\sigma x^\nu
-\nonumber \\
&-&\tau_F^2\partial_\sigma x^\sigma \tau_\sigma^{ \ b}\epsilon_{ba}m_\mu^{ \ a}
h^{\mu\nu}m_\nu^{ \ c}\epsilon_{cd}\tau_\rho^{ \ d}\partial_\sigma x^\rho+
\tau_F^2\partial_\sigma x^\mu h_{\mu\nu}\partial_\sigma x^\nu
\nonumber \\
&\equiv &p_\mu h^{\mu\nu}p_\nu+p_\mu V^\mu+\tau_F^2\partial_\sigma x^\mu \bar{H}_{\mu\nu}\partial_\sigma x^\nu \ , \quad V^\mu=V^\mu_{ \ \nu}\partial_\sigma x^\nu \ ,  \nonumber \\
\end{eqnarray}
where  we identify $\ttau_F$ with $\tau_F$  since as follows from the analysis above  it is not necessary
to rescale $\tau_F$ in order to have finite Hamiltonian in the limit $\omega\rightarrow \infty$.

We see that this form of the Hamiltonian constraint is rather complicated. For that reason it is necessary to check  whether it defines consistent theory. Especially we would like to see whether  Hamiltonian and  spatial diffeomorphism constraints are the first class constraints.
To do this we
calculate  Poisson algebra of constraints. As usually we introduce smeared form of these constraints
 \begin{eqnarray}
 \bT_\tau(N)=\int d\sigma N \mH_\tau \ , \quad
 \bT_\sigma(N^\sigma)=\int d\sigma N^\sigma \mH_\sigma
 \nonumber \\
 \end{eqnarray}
 and we easily find
 \begin{eqnarray}\label{pbbSbS}
 \pb{\bT_\sigma(N^\sigma),\bT_\sigma(M^\sigma)}=
 \int d\sigma (N^\sigma\partial_\sigma M^\sigma-N^\sigma
 \partial_\sigma M^\sigma)p_\mu\partial_\sigma x^\mu=
 \bT_\sigma(N^\sigma\partial_\sigma M^\sigma-N^\sigma
 \partial_\sigma M^\sigma) \ .  \nonumber \\
 \end{eqnarray}
 In case of the calculation of the Poisson brackets of two  Hamiltonian constraints the situation is more involved since the explicit calculation gives
 \begin{eqnarray}
 & & \pb{\bT_\tau(N),\bT_\tau(M)}=\int d\sigma
 (N\partial_\sigma M-M\partial_\sigma N)2\tau_F^2(p_\mu h^{
    \mu\nu}\bar{H}_{\nu\rho}\partial_\sigma x^\rho+
 \partial_\sigma x^\rho \bar{H}_{\rho \mu}h^{\mu\nu}p_\nu)
 +
 \nonumber \\
 &-&2\int d\sigma \tau_F (N\partial_\sigma M-M\partial_\sigma N)
 p_\mu V^\mu_{ \ \nu}h^{\nu\omega}p_\omega+
 \nonumber \\
 &+&\int d\sigma (N\partial_\sigma M-M\partial_\sigma N)
 p_\rho V^\rho_{ \ \sigma}V^\sigma_{\ \omega}\partial_\sigma x^\omega+
 \nonumber \\
 &-&\tau_F^2\int d\sigma (N\partial_\sigma M-M\partial_\sigma N)(V^\mu_{ \ \nu}\partial_\sigma x^\nu \bar{H}_{\mu\rho}\partial_\sigma x^\rho+
 \partial_\sigma x^\rho \bar{H}_{\rho\mu}V^\mu_{ \ \nu}\partial_\sigma x^\nu)  \ .
\nonumber \\
 \end{eqnarray}
 To proceed further we calculate
 \begin{eqnarray}
& & 2p_\mu h^{\mu\nu}\bar{H}_{\nu\rho}\partial_\sigma x^\rho+2\partial_\sigma x^\rho
 \bar{H}_{\rho\mu}h^{\mu\nu}p_\nu= \nonumber \\
 & &=
 4\tau_F^2p_\mu h^{\mu\nu}h_{\nu\rho}\partial_\sigma x^\rho+
 4\tau_F^2 \partial_\sigma x^\mu \tau_\mu^{ \ a}\eta_{ab}m_\nu^{ \ b}
 h^{\nu\rho}p_\rho \ ,
 \nonumber \\
& & p_\rho V^\rho_{ \ \mu}V^\mu_{ \ \nu}\partial_\sigma x^\nu
= 4\tau_F^2 p_\mu \tau^{\mu\nu}\tau_{\nu\rho}\partial_\sigma x^\rho-
 4\tau_F^2p_\mu h^{\mu\nu}m_\nu^{ \ a}\tau_\rho^{ \ b}\eta_{ab}\partial_\sigma x^\rho
\ ,  \nonumber \\
& & V^\mu_{ \ \nu}\partial_\sigma x^\nu \bar{H}_{\mu\rho}\partial_\sigma x^\rho+
 \partial_\sigma x^\rho \bar{H}_{\rho\mu}V^\mu_{ \ \nu}\partial_\sigma x^\nu=0 \ ,
 \quad  p_\mu V^\mu_{ \ \nu}h^{\nu\omega}p_\omega=0 \ . \nonumber \\
  \end{eqnarray}
  Collecting these results together we finally obtain
 \begin{equation}\label{pbbTbT}
 \pb{\bT_\tau(N),\bT_\tau(M)}=\bT_\sigma ((N\partial_\sigma M-M\partial_\sigma N)
 4\tau_F^2 ) \
 \end{equation}
 which is the correct form of the Poisson bracket between Hamiltonian constraints.
 Finally we calculate the Poisson bracket
 \begin{equation}
 \pb{\bT_\sigma(N^\sigma),\bT_\tau(M)} \ .
 \end{equation}
 Since
 \begin{eqnarray}
 \pb{\bT_\sigma(N^\sigma),x^\mu}=-N^\sigma\partial_\sigma x^\mu \ ,  \quad
 \pb{\bT_\sigma(N^\sigma),p_\mu}=-\partial_\sigma N^\sigma p_\mu-N^\sigma \partial_\sigma p_\mu
 \nonumber \\
 \end{eqnarray}
 we easily find
 \begin{equation}
 \pb{\bT_\sigma (N^\sigma),\mH_\tau}=-2\partial_\sigma N^\sigma \mH_\tau-
 N^\sigma \partial_\sigma \mH_\tau
 \end{equation}
 or alternatively
 \begin{equation}\label{pbbTbS}
 \pb{\bT_\sigma (N^\sigma),\bT_\tau(M)}=\bT_\sigma (N^\sigma\partial_\sigma M-
 \partial_\sigma N^\sigma M) \ .
 \end{equation}
 We see that all Poisson brackets (\ref{pbbSbS}),(\ref{pbbTbT}) and
 (\ref{pbbTbS}) vanish  on the constraint surface
 $\mH_\tau\approx 0 ,\mH_\sigma \approx 0 $ and hence they are the first class constraints and the non-relativistic string is well defined system from the canonical point of view.
\section{Lagrangian Form}
\label{third}
In this section we focus on the Lagrangian formulation of the proposed Hamiltonian form of non-relativistic string in Newton-Cartan background. Recall that this string is defined with the Hamiltonian constraint (\ref{mHtau}) and the spatial diffeomorphism constraint $\mH_\sigma\approx 0$.   In order to understand subtle points in the
transformation from the Hamiltonian to Lagrangian description of this system  we firstly start with the simpler problem of non-relativistic string in the flat background.
\subsection{Flat space-time limit}
The non-relativistic string in the flat background has  following Hamiltonian
\begin{equation}\label{Hamflat}
H=\int d\sigma (\lambda^\tau \mH_\tau+\lambda^\sigma \mH_\sigma) \ ,
\end{equation}
where
\begin{equation}
\mH_\tau=-2\tau_F p_a \eta^{ab}e_{bc}\partial_\sigma x^c+
p_i h^{ij} p_j+\tau^2_F h_{ij}\partial_\sigma x^i\partial_\sigma x^j \ , \quad
\mH_\sigma=p_i \partial_\sigma x^i +p_a\partial_\sigma x^a \ ,
\end{equation}
where $a,b,c,\dots=0,1$ and where $h_{ij}=\delta_{ij} \ , h^{ij}=\delta^{ij} \ , i,j,\dots=2,\dots,d-1$.
Our goal is to find Lagrangian formulation of the non-relativistic string
in flat background. With the help of the Hamiltonian
 (\ref{Hamflat}) we obtain following equations of motion for $x^0,x^1$ and $x^i$
\begin{eqnarray}\label{eqflat}
\partial_\tau x^0&=&\pb{x^0,H}=-2\tau_F\lambda^\tau \partial_\sigma x^1+\lambda^\sigma \partial_\sigma x^0 \ , \nonumber \\
\partial_\tau x^1&=&\pb{x^1,H}=-2\tau_F\lambda^\tau \partial_\sigma x^0+\lambda^\sigma
\partial_\sigma x^1 \ , \nonumber \\
\partial_\tau x^i&=&\pb{x^i,H}=2\lambda^\tau h^{ij}p_j+\lambda^\sigma \partial_\sigma x^i \ .
\nonumber \\
\end{eqnarray}
Then it is easy to find corresponding Lagrangian density
\begin{eqnarray}\label{Lagdenflat}
& &\mL=p_a\partial_\tau x^a+p_i\partial_\tau x^i-\mL=\lambda^\tau p_i h^{ij}p_j-\lambda^\tau \partial_\sigma x^i\partial_\sigma x^j h_{ij}=\nonumber \\
& &=\frac{1}{4\lambda^\tau}(\partial_\tau x^i-\lambda^\sigma \partial_\sigma x^i)h_{ij}
(\partial_\tau x^j-\lambda^\sigma \partial_\sigma x^j)-\lambda^\tau \tau^2_F h_{ij}
\partial_\sigma x^i\partial_\sigma x^j \ .  \nonumber \\
\end{eqnarray}
We see that this Lagrangian does not depend on the variables $x^a$ which is confusing   since if we  perform inverse Legendre transformation from
(\ref{Lagdenflat}) and determine corresponding Hamiltonian we will find that it does not depend on $p_a$. We can resolve this problem when we closely examine equations of motion for $x^0$ and $x^1$. We firstly  consider the
first equation in (\ref{eqflat}) and
multiply it with  $\partial_\sigma x^0$ while we multiply the second one with  $\partial_\sigma x^1$. Then if we take their   difference we  obtain
\begin{equation}
-\partial_\tau x^0\partial_\sigma x^0+\partial_\tau x^1\partial_\sigma x^1=
\lambda^\sigma (\partial_\sigma x^1\partial_\sigma x^1-\partial_\sigma x^0\partial_\sigma x^0)
\end{equation}
that can be solved for $\lambda^\sigma$ as
\begin{equation}\label{lambdasigmasol}
\lambda^\sigma=\frac{\ba_{\tau\sigma}}{\ba_{\sigma\sigma}} \ , \quad \ba_{\alpha\beta}=
\partial_\alpha x^a\partial_\beta x^b\eta_{ab} \ .
\nonumber \\
\end{equation}
On the other hand from the equations of motion for $x^0$ and $x^1$ we obtain
\begin{eqnarray}
(\partial_\tau x^0-\lambda^\sigma \partial_\sigma x^0)^2=4(\lambda^\tau)^2\partial_\sigma x^1\partial_\sigma x^1 \ , \quad
(\partial_\tau x^1-\lambda^\sigma \partial_\sigma x^1)^2=4(\lambda^\tau)^2\partial_\sigma x^0\partial_\sigma x^0 \  \nonumber \\
\end{eqnarray}
that implies
\begin{equation}
-\ba_{\tau\tau}+2\lambda^\sigma \ba_{\sigma\tau}-(\lambda^\sigma)^2\ba_{\sigma\sigma}=
4(\lambda^\tau)^2\ba_{\sigma\sigma} \tau^2_F \ .
\end{equation}
Inserting (\ref{lambdasigmasol}) into this equation  we find that $\lambda^\tau$ is equal to
\begin{equation}\label{lambdatausol}
\lambda^\tau=
\frac{\sqrt{-\det \ba_{\alpha\beta}}}{2\tau_F \ba_{\sigma\sigma}} \ .
\end{equation}
Then if we combine (\ref{lambdasigmasol}) together with (\ref{lambdatausol})
we get
\begin{eqnarray}\label{lambdafinal}
& &\frac{1}{\lambda^\tau}=-2\tau_F \ba^{\tau\tau}\sqrt{-\det\ba} \ , \quad
\frac{2\lambda^\sigma}{\lambda^\tau}=4\tau_F \ba^{\tau\sigma}\sqrt{-\det \ba}
\ , \nonumber \\
& &\frac{(\lambda^\sigma)^2}{4(\lambda^\tau)^2}-\lambda^\tau \tau^2_F=-\tau_F \ba^{\sigma\sigma}\sqrt{-\det\ba} \ .
\end{eqnarray}
Finally  inserting (\ref{lambdafinal}) into
(\ref{Lagdenflat}) we obtain
\begin{eqnarray}
\mL
=-\frac{\tau_F}{2}\sqrt{-\det \ba}\ba^{\alpha\beta}h_{\alpha\beta} \
\nonumber \\
\end{eqnarray}
which has exactly  the same form as  the Lagrangian density that was derived in \cite{Andringa:2012uz}.

\subsection{Lagrangian for String in Newton-Cartan Background}
 Now we proceed to the case of non-relativistic string in Newton-Cartan background. As the first step we formulate the Hamiltonian constraint
 with the help of the variables that reflect an invariance of the theory under generalized Galilean boosts that have the form  \cite{Andringa:2012uz}
 \begin{equation}\label{Galltr}
 \delta e_\mu^{ \ a'}=\tau_\mu^{ \ a}\lambda_a^{ \ a'} \ ,
 \quad
 \delta \tau^\mu_{\ a}=e^\mu_{ \ a'}\lambda^{a'}_{ \ a} \ , \quad \delta m_\mu^{ \ a}=
 e_\mu^{ \ a'}\lambda_{a'}^{ \ a} \ ,
 \end{equation}
 where $\lambda_a^{ \ a'}$ are parameters that obey following relations
 \begin{eqnarray}\label{Galltrpar}
& & \eta_{ac}\lambda^c_{ \ a'}+
 \delta_{a'b'}\lambda^{b'}_{ \ a}=0 \ , \quad \lambda_{a'}^{ \ c}\eta_{ca}+
 \lambda_a^{ \ b'}\delta_{b'a'}=0 \ ,
 \nonumber \\
& &  \lambda_{a'}^{ \ a}+\lambda^{a}_{ \ a'}=0 \ ,  \quad  \lambda^{a'}_{ \ a}+\lambda_a^{ \ a'}=0 \ .
 \nonumber \\
 \end{eqnarray}
 Now we define boost invariant temporary vierbein as $\hat{\tau}^\mu_{ \ a}$
\begin{equation}
\hat{\tau}^\mu_{ \ a}=\tau^\mu_{ \ a}-h^{\mu\nu}m_\nu^{ \ b}\eta_{ba}  \ .
\end{equation}
This is invariant under (\ref{Galltr}) since
\begin{equation}
\delta \hat{\tau}_\mu^{ \ a}=
e^\mu_{ \ a'}\lambda^{a'}_{ \ a}-e^{\mu}_{ \ c'}\delta^{c'b'}\lambda_{b'}^{ \ b}\eta_{ba}=
e^\mu_{ \ a'}\lambda^{ a'}_{ \ a}+e^\mu_{ \ a'}\lambda_a^{ \ a'}=0 \ ,
\end{equation}
where in the last step we used (\ref{Galltrpar}). With the help of  $\hat{\tau}_\mu^{ \ a}$ we can rewrite $V^\mu$ into manifestly invariant form
\begin{eqnarray}
V^\mu=
V^\mu_{ \ \nu}\partial_\sigma x^\nu \ ,
V^\mu_{ \ \nu}=
-2\tau_F \hat{\tau}^\mu_{ \ a}\epsilon^{ab}
\hat{\tau}^\sigma_{ \ b}\tau_{\sigma\nu}=V^{\mu\sigma}\tau_{\sigma\nu} \ , \nonumber \\
\end{eqnarray}
where $\epsilon^{ab}\equiv \eta^{ac}\eta^{bd}\epsilon_{cd}$ and where
$V^{\mu\nu}=-V^{\nu\mu}$.
Let us now analyze in more details the object $\bar{H}_{\mu\nu}$. After some calculations we obtain that it can be rewritten into the form
\begin{eqnarray}\label{bHnew}
\bar{H}_{\mu\nu}&=&-\tau_F^2 \tau_\mu^{ \ c}\epsilon_{cd}\Phi^{da}\epsilon_{ab}
\tau_\nu^{ \ b}+\tau_F^2\bh_{\mu\nu} \ ,  \nonumber \\
\bh_{\mu\nu}&=&h_{\mu\nu}+m_\mu^{ \ a}\tau_\nu^{ \ b}\eta_{ab}+
\tau_\mu^{ \ a}m_\nu^{ \ b}\eta_{ab} \  , \nonumber \\
\Phi^{ab}&=&-\tau^{\mu}_{ \ d}\eta^{da}m_\mu^{ \ b}-
m_\mu^{ \ a}\tau^{\mu}_{ \ d}\eta^{db}+m_\mu^{ \ a}h^{\mu\nu}
m_\nu^{ \ b} \ .  \nonumber \\
\end{eqnarray}
An important property of (\ref{bHnew}) is that it is written
with the help of the objects that are invariant under
(\ref{Galltr}). To see this let us firstly consider variation of
$\bh_{\mu\nu}$
\begin{eqnarray}
& &\delta \bh_{\mu\nu}=\tau_\mu^{\ a}\lambda_a^{ \ a'}\delta_{a'b'}
e_\nu^{ \ b'}+e_\mu^{ \ a'}\delta_{a'b'}\tau_\nu^{ \ b}\lambda_b^{ \ b'}+
\nonumber \\
& &+e_\mu^{ \ a'}\lambda_{a'}^{ \ a}\tau_\nu^{ \ b}\eta_{ab}+
\tau_\mu^{ \ a}\eta_{ab}e_\nu^{ \ b'}\lambda_{b'}^{ \ b}=0 \ ,
\nonumber \\
\end{eqnarray}
where we used (\ref{Galltrpar}). Finally we consider the variation of
$\Phi^{ab}=\Phi^{ba}$. Note that the matrix $\Phi^{ab}$ can be interpreted as the matrix of Newton-potential in generalized Newton-Cartan gravity which  is now
matrix valued as opposite to the scalar form of $\Phi$ in ordinary Newton-Cartan gravity. On the other hand it is still invariant under
(\ref{Galltr}) since
\begin{eqnarray}
\delta \Phi^{ab}=
-e^\mu_{ \ a'}\lambda^{a'}_{ \ c}m_\mu^{ \ b}\eta^{ca}-
m_\mu^{ \ a}e^{\mu}_{ \ a'}\lambda^{a'}_{ \ d}\eta^{db}
+e^\nu_{ \ d'}\delta^{c'a'}\lambda_{a'}^{ \ a}m_\nu^{ \ b}+
m_\mu^{ \ a}e^\mu_{ \ c'}\delta^{c'b'}\lambda_{b'}^{ \ b}=
\nonumber \\
=-e^\mu_{ \ a'}\lambda^{a'}_{ \ c}m_\mu^{ \ b}\eta^{ca}-
m_\mu^{ \ a}e^{\mu}_{ \ a'}\lambda^{a'}_{ \ d}\eta^{db}
-e^\nu_{ \ c'}\delta^{c'a'}\lambda_{ \  a'}^{a}m_\nu^{ \ b}-
m_\mu^{ \ a}e^\mu_{ \ c'}\delta^{c'b'}\lambda_{ \ b'}^{b}=
\nonumber \\
=-e^\mu_{ \ a'}\lambda^{a'}_{ \ c}m_\mu^{ \ b}\eta^{ca}-
m_\mu^{ \ a}e^{\mu}_{ \ a'}\lambda^{a'}_{ \ d}\eta^{db}+
e^\mu_{ \ c'}\lambda^{c'}_{ \ c}\eta^{ca}m_\mu^{ \ b}
+m_\mu^{ \ a}e^\mu_{ \ c'}\lambda^{c'}_{ \ c}\eta^{cb}=0 \ ,
\nonumber \\
\end{eqnarray}
where in the first step we used the relation $\lambda_{a'}^{ \ a}=-\lambda^a_{ \ a'}$ and in the last step we used the fact that
\begin{equation}
\lambda^{c'}_{ \ b}\eta^{ba}+\lambda^c_{ \ b'}\delta^{b'c'}=0 \ .
\end{equation}
It is also instructive to elaborate more about an expression that contain the potential $\Phi^{ab}$. We find that after some calculations it can be rewritten into the form
\begin{eqnarray}
&- &\tau_F^2 \partial_\sigma x^\mu \tau_\mu^{ \ c} \epsilon_{cd}\Phi^{da}\epsilon_{ab}\tau^b_{ \ \nu}\partial_\sigma x^\nu=\nonumber \\
&=&\tau_F^2\partial_\sigma x^\mu\tau_\mu^{ \ c}\eta_{ca}\Phi^{ab}\eta_{bd}\tau^{ \ d}_\nu\partial_\sigma x^\nu-\tau_F^2 \ba_{\sigma\sigma}
\Phi^{ab}\eta_{ba} \nonumber \\
\end{eqnarray}
so that we finally obtain manifestly invariant Hamiltonian in the form
\begin{eqnarray}\label{Hinvariant}
& &H=\int d\sigma (\lambda^\tau \mH_\tau+\lambda^\sigma \mH_\sigma) \ , \quad \mH_\sigma=p_\mu\partial_\sigma x^\mu \ , \nonumber \\
& &\mH_\tau=p_\mu h^{\mu\nu}p_\nu-2p_\mu \tau_F \hat{\tau}^\mu_{ \ a}\epsilon^{ab}
\hat{\tau}^\sigma_{ \ b}\tau_{\sigma\nu}\partial_\sigma x^\nu
+\nonumber \\
& &+\tau_F^2\tau^c\eta_{ca}\Phi^{ab}\eta_{bd}\tau^d-\tau_F^2 \ba_{\sigma\sigma}
\Phi^{ab}\eta_{ba}+\tau_F^2\bh_{\sigma\sigma}  \ , \nonumber \\
\end{eqnarray}
where $\tau^a=\partial_\sigma x^\mu \tau_\mu^{ \ a},\bh_{\sigma\sigma}=
\partial_\sigma x^\mu \bh_{\mu\nu}\partial_\sigma x^\nu$ and where $\lambda^\tau$ and
$\lambda^\sigma$ are corresponding Lagrange multipliers.

Before we proceed to the Lagrangian formulation of the theory let us introduce
 vierbein $\hat{e}_\mu^{ \ a'}$ defined
as
\begin{equation}\label{hate}
\hat{e}_\mu^{ \ a'}=e_\mu^{ \ a'}+m_\nu^{ \ a}e^\nu_{ \ c'}\delta^{c'a'}
\tau_\mu^{ \ b}\eta_{ba} \ ,
\end{equation}
that is again invariant under (\ref{Galltr})
\begin{eqnarray}
\delta \hat{e}_\mu^{ \ a'}
=\tau_\mu^{ \ a}\lambda_a^{ \ a'}+\lambda_{c'}^{ \ a}\delta^{c'a'}
\tau_\mu^{ \ b}\eta_{ba}=\nonumber \\
=\tau_\mu^{ \ a}\lambda_a^{ \ a'}
-\lambda_b^{ \ b'}\delta_{b'c'}\delta^{c'a'}\tau_\mu^{ \ b}=0 \ . \nonumber \\
\end{eqnarray}
Note that we have following useful identity
\begin{eqnarray}
\hat{\tau}^\mu_{ \ a}\hat{e}_\mu^{ \ a'}=0
\end{eqnarray}
and also
\begin{equation}
\hat{e}_\mu^{ \ a'}e^\mu_{ \ b'}=\delta^{ a'}_{ \ b'} \ , \quad
\hat{e}_\mu^{ \ a'}h^{\mu\nu}=e^\nu_{ \ c'}\delta^{c'a'}  \ .
\end{equation}
 Now we are ready to proceed to the Lagrangian formulation of the theory. We begin with the canonical equations of motion for $x^\mu$ that follow from the
 Hamiltonian (\ref{Hinvariant})
\begin{equation}\label{eqmXmu}
\partial_\tau x^\mu=\lambda^\tau (2h^{\mu\nu}p_\nu+V^\mu)+\lambda^\sigma \partial_\sigma x^\nu \ .
\end{equation}
Let us now  multiply this equation  with $\hat{e}_\mu^{ \ a'}$. Using the fact that
$\hat{e}_\mu^{ \ a'}\hat{\tau}^\mu_{ \ b'}=0$ we find that
$\hat{e}_\mu^{ \  a'}V^\mu=0$ and from (\ref{eqmXmu}) we obtain
\begin{equation}
\hat{e}_\mu^{ \ a'}\partial_\tau x^\mu=2\lambda^\tau e^\mu_{ \ c'}\delta^{c'a'}
p_\mu+\lambda^\sigma \hat{e}_\mu^{ \ a'}\partial_\sigma x^\mu \
\end{equation}
and consequently
\begin{eqnarray}
(\partial_\tau x^\mu-\lambda^\sigma\partial_\sigma x^\mu)\hat{e}_\mu^{ \ a'}\delta_{a'b'}
\hat{e}_\nu^{ \ b'}(\partial_\tau x^\nu-\lambda^\sigma \partial_\sigma x^\nu)
=4(\lambda^\tau)^2p_\mu h^{\mu\nu}p_\nu  \ . \nonumber \\
\end{eqnarray}
With the help of this result we easily find the Lagrangian density in the form
\begin{eqnarray}\label{mLHam}
\mL&=&p_\mu\partial_\tau x^\mu-\lambda^\tau \mH_\tau-\lambda^\sigma \mH_\sigma=
\nonumber \\
&=&\frac{1}{4\lambda^\tau}(\partial_\tau x^\mu-\lambda^\sigma\partial_\sigma x^\mu)\hat{e}_\mu^{ \ a'}\delta_{a'b'}
\hat{e}_\nu^{ \ b'}(\partial_\tau x^\nu-\lambda^\sigma \partial_\sigma x^\nu)-
\tau_F^2 \lambda^\tau \bar{H}_{\sigma\sigma} \ . \nonumber \\
\end{eqnarray}
To proceed further we observe that (\ref{eqmXmu})
implies
\begin{eqnarray}
& & \hat{e}_\mu^{ \ a'}\delta_{a'b'}\hat{e}_\nu^{ \ b'}
=\bh_{\mu\nu}+\tau_\mu^{ \ c}\eta_{ca}\Phi^{ab}\eta_{bd}\tau_\nu^{ \ d} \ ,
\nonumber \\
\end{eqnarray}
where we also used
\begin{equation}
e_\mu^{ \ a'}e^\nu_{ \ a'}=\delta_\mu^\nu-\tau_\mu^{ \ a}\tau^\nu_{ \ a} \ .
\end{equation}
Then we can rewrite the Lagrangian density
(\ref{mLHam}) into the form
\begin{eqnarray}\label{mLHam1}
& &\mL=\frac{1}{4\lambda^\tau}(\bh_{\tau\tau}-2\lambda^\sigma
\bh_{\sigma\tau}+(\lambda^\sigma)^2\bh_{\sigma\sigma}+\nonumber \\
&+&\partial_\tau x^\mu \tau_\mu^{ \ c}\eta_{ca}\Phi^{ab}\eta_{bd}
\tau_\nu^{ \ d}\partial_\tau x^\nu-2\lambda^\sigma
\partial_\tau x^\mu \tau_\mu^{ \ c}\eta_{ca}\Phi^{ab}\eta_{bd}\tau_\nu^{ \ d}
\partial_\sigma x^\nu+(\lambda^\sigma)^2
\partial_\sigma x^\mu \tau_\mu^{ \ c}\eta_{ca}\Phi^{ab}\eta_{bd}\tau_\nu^{ \ d}\partial_\sigma x^\nu)-\nonumber \\
&-&\lambda^\tau\tau_F^2\partial_\sigma x^\mu\tau_\mu^{ \ c}\eta_{ca}\Phi^{ab}\eta_{bd}\tau_\nu^{ \ d}\partial_\sigma x^\nu+\lambda^\tau
\tau_F^2 \ba_{\sigma\sigma}
\Phi^{ab}\eta_{ba}-\lambda^\tau\tau_F^2\bh_{\sigma\sigma} \ , \nonumber \\
\end{eqnarray}
where $\bh_{\alpha\beta}=\bh_{\mu\nu}\partial_\alpha x^\mu \partial_\beta x^\nu$.

Finally we eliminate  $\lambda^\tau$ and $
\lambda^\sigma$ from (\ref{mLHam1}).  As in case of the flat space-time limit their form is not determined
by their equations of motion. Instead they can be determined using the equations of motion for $x^\mu$. In fact, if we multiply (\ref{eqmXmu}) by $\tau_{\mu\nu}$ and use
the fact that $\tau_{\mu\nu}h^{\nu\rho}=0$ we obtain
\begin{equation}
\tau_{\mu\nu}(\partial_\tau x^\nu-\lambda^\sigma \partial_\sigma x^\nu)
-\lambda^\tau \tau_{\mu\nu}V^\nu=0 \ .
\end{equation}
We can multiply this equation with $\partial_\sigma x^\mu$ and we obtain
\begin{equation}
\lambda^\sigma=\frac{\ba_{\sigma\tau}}{\ba_{\sigma\sigma}} \ , \quad  \ba_{\alpha\beta}=
\partial_\alpha x^\mu \tau_{\mu\nu}\partial_\beta x^\nu
\end{equation}
using the fact that
\begin{equation}
\partial_\sigma x^\mu \tau_{\mu\nu}V^\nu=
2\tau_F\partial_\sigma x^\mu \tau_\mu^{ \ a}\epsilon_{ab}\tau_\nu^{ \  b}\partial_\sigma x^\nu=0 \ .
\end{equation}
In the similar way we obtain
\begin{eqnarray}
(\partial_\tau x^\mu -\lambda^\sigma \partial_\sigma x^\mu)\tau_{\mu\nu}
(\partial_\tau x^\nu-\lambda^\sigma \partial_\sigma x^\nu)=(\lambda^\tau)^2V^\mu \tau_{\mu\nu}V^\nu \nonumber \\
\end{eqnarray}
that can be solved for $\lambda^\tau$ as
\begin{equation}
\lambda^\tau=\frac{\sqrt{-\det\ba_{\alpha\beta}}}{\sqrt{-V^\mu\tau_{\mu\nu}V^\nu}\sqrt{\ba_{\sigma\sigma}}} \ ,
\nonumber \\
\end{equation}
where
\begin{eqnarray}
V^\mu \tau_{\mu\nu}V^\nu
=-4\tau_F^2\ba_{\sigma\sigma} \ .
\nonumber \\
\end{eqnarray}
Now we see that we can proceed as in the case of the non-relativistic
string in flat space-time and we obtain the final result
\begin{eqnarray}\label{mLfin}
\mL=-\frac{\tau_F}{2}
\sqrt{-\det\ba}\left(\ba^{\alpha\beta}\bh_{\alpha\beta}
+\ba^{\alpha\beta}\partial_\alpha x^\mu \tau_\mu^{ \ c}\eta_{ca}\Phi^{ab}
\eta_{bd}\tau_\nu^{ \ d}\partial_\beta x^\nu-\Phi^{ab}\eta_{ba}\right)
\nonumber \\
\end{eqnarray}
We see that this Lagrangian density almost coincides with the Lagrangian density
found \cite{Andringa:2012uz} up to terms that contain matrix valued Newton potential
$\Phi_{ab}$. Now we are going to argue that these terms cancel each other. In fact,
note that $\ba_{\alpha\beta}$ is defined as
\begin{equation}
\ba_{\alpha\beta}=\tau_\alpha^{ \ a}\tau_\beta^{ \ b}\eta_{ab} \ ,
\end{equation}
where $\tau_{\alpha}^{ \ a}\equiv \partial_\alpha x^\mu \tau_\mu^{ \ a}$ is $2\times 2$ matrix. Since 
 $\tau_{\alpha\beta}$ is
non-singular so that $\tau_{\alpha}^{ \ a}$  is non-singular as well
due to the fact that 
   \begin{equation}
   \det \ba_{\alpha\beta}=(\det \tau_\alpha^{ \ a})^2\det \eta_{ab}
   =-(\det\tau_\alpha^{ \ a})^2\neq 0 \ .
   \end{equation}
   Then  we can introduce
an inverse matrix $\tau^\beta_{\ a}$ that obeys
the relation
\begin{equation}
\tau^\alpha_{ \ a}\tau_\alpha^{ \ b}=\delta_a^{ \ b} \ .
\end{equation}
As a result we can
define $\ba^{\alpha\beta}$ as
\begin{equation}
\ba^{\alpha\beta}=\tau^\alpha_{ \ a}\tau^\beta_{ \ b}\eta^{ab}
\end{equation}
that obeys
\begin{equation}
\ba^{\alpha\beta}\tau_\beta^{ \ a}=\tau^\alpha_{ \ c}\eta^{ca} \ ,
\end{equation}
and hence
\begin{equation}
\ba^{\alpha\beta}\tau_\beta^{ \ b}\tau_\alpha^{ \ a}=
\tau_\beta^{ \ b}\tau^\beta_{ \ c}\eta^{ca}=\eta^{ba} \ .
\end{equation}
With the help of these results we can manipulate with the second term in 
(\ref{mLfin}) as
\begin{eqnarray}
-\frac{\tau_F}{2}\sqrt{-\det \ba}\ba^{\alpha\beta}
\tau_\alpha^{ \ a}\Phi_{ab}\tau_\beta^{ \ b}=-\frac{\tau_F}{2}
\sqrt{-\det\ba}\eta^{ab}\Phi_{ba}
\nonumber \\
\end{eqnarray}
and we see that it exactly cancels the last term in (\ref{mLfin}). As
a result we derive 
 the Lagrangian density in the  final form
\begin{equation}
\mL=-\frac{T}{2}\sqrt{-\det\ba}\ba^{\alpha\beta}
\bh_{\alpha\beta}
\end{equation}
which is Lagrangian density proposed in \cite{Andringa:2012uz}. This result
again confirms validity of our approach.

\section{Conclusion}\label{fourth}
Let us outline our results and suggest possible extension of this work. We found Hamiltonian for non-relativistic string in Newton-Cartan background from the Hamiltonian of relativistic string in general background when we used the limiting procedure introduced in
\cite{Bergshoeff:2015uaa}. The corresponding Hamiltonian is linear combination of two constraints and we checked that they are the first class constraints which is a consequence of diffeomorphism invariance of world-sheet theory. We also introduced variables that are invariant under Milne boost and we showed that the Hamiltonian constraint is invariant under this transformation too. Finally we found Lagrangian formulation of the non-relativistic string in Newton-Cartan background that agrees with the Lagrangian density proposed in
\cite{Andringa:2012uz}. We mean that this is very nice consistency check of our result.

This paper can be extended in different directions. It would be possible to perform similar analysis in  the case of non-relativistic p-brane in Newton-Cartan background. Secondly, we could also extend this analysis to the case of superstring.
We hope to return to some of these problems in future.
\\
\\
{\bf Acknowledgment:}

This  work  was
    supported by the Grant Agency of the Czech Republic under the grant
    P201/12/G028.

\end{document}